\documentstyle[12pt]{article}
\def\a{\alpha}
\def\b{\beta}
\def\c{\chi}
\def\d{\delta}
\def\e{\epsilon}

\def\g{\gamma}

\def\p{\psi}

\def\l{\lambda}
\def\m{\mu}

\def\o{\omega}

\def\s{\sigma}

\def\vf{\varphi}

\def\bz{\bar{z}}

\def\bp{\bar{\p}}

\def\be{\begin{equation}}
\def\ee{\end{equation}}
\def\arr{\begin{array}{rll}}
\def\ea{\end{array}}
\def\bea{\begin{eqnarray}}
\def\eea{\end{eqnarray}}

\begin{document}

\begin{titlepage}
\noindent

\vskip 1.5cm

\begin{center}

{\Large\bf Restoring Lorentz Invariance }\\

\vskip 0.5cm

{\Large\bf in Classical N=2 String }\\

\bigskip

\vskip 1.5cm

{\large Stefano Bellucci}~\footnote{
E-mail: Stefano.Bellucci@lnf.infn.it} 
and 
{\large Anton Galajinsky}~\footnote{
On leave from Department of Mathematical Physics,
Tomsk Polytechnical University, Tomsk, Russia\\
\phantom{XX}   
E-mail: Anton.Galajinsky@lnf.infn.it}

\vskip 0.4cm

{\it  INFN--Laboratori Nazionali di Frascati, C.P. 13, 
00044 Frascati, Italia}
\vskip 0.4cm

\end{center}

\vskip 1.5cm

\begin{abstract}
\noindent
We study classical N=2 string within the framework of the 
N=4 topological formalism by Berkovits and Vafa. Special emphasis is 
put on the demonstration of a classical equivalence of the theories 
and the construction of an action for the N=4 topological string. 
The $SO(2,2)$ Lorentz invariance missing in the conventional Brink--Schwarz 
action for the $N=2$ string is restored in the $N=4$ topological action.
\end{abstract}

\vspace{0.5cm}

PACS: 04.60.Ds; 11.30.Pb\\ \indent
Keywords: $N{=}2$ string, Lorentz symmetry

\end{titlepage}

\noindent
{\bf 1. Introduction}\\[-4pt]

\vskip 0.4cm

Nowadays, it seems to be a conventional wisdom to view string theories
revealing an $N$--extended local supersymmetry on the world--sheet as 
models describing a coupling of an $N$--extended (conformal) $d=2$ 
supergravity to matter multiplets. Technically, it suffices to start 
with an appropriate supergravity multiplet, the component fields being 
connected by a set of gauge symmetries, and require the coupling not 
to destroy the transformations. There has been considerable interest
in the field~\cite{brink}--\cite{beliv} which culminated in the 
classification of the underlying $N$--extended superconformal 
algebras~\cite{ramond}--\cite{schoutens1}. The $N\le 4$ bound was encountered
as the only one compatible with the presence of a central charge in 
the superconformal algebra (SCA). It should be stressed that with $N$ growing 
the construction of an action becomes technically involved.
For example the $N=4$ action of Ref.~\cite{pernpvn} appeals first to 
a dimensional reduction of the rigid $N=2$, $d=4$ nonlinear sigma model 
to $d=2$ which is then coupled to a supergravity multiplet. 
Notice also 
that for $N > 1$ the existing actions~\cite{brink,pernpvn} lack manifest 
Lorentz invariance in a target.

The $N=2$ case, which is also the subject of this paper, is of particular 
interest. In contrast to $N=0$ and $N=1$ strings, quantization of the model 
reveals only a finite number of physical states in the spectrum. What is more,
one encounters a continuous family of sectors interpolating between $R$ and 
$NS$ which are connected by spectral flow~\cite{schwimmer}. The latter 
fact, in particular, allows one to stick with a preferred sector. All 
tree--level amplitudes with more than three external legs prove to 
vanish~\cite{ov1, ov} (for details of loop calculations  see a 
recent work~\cite{cln} and references therein), exhibiting a 
remarkable connection with self--dual gauge or gravitational theory in 
two spatial and two temporal dimensions~\cite{ov}. Turning to problematic
points, in spite of being a theory of an $N=2$ $d=2$ supergravity coupled to 
matter, there are no space-time fermions in the quantum spectrum. 
Furthermore, as has been mentioned above manifest Lorentz covariance is 
missing in the Brink--Schwarz action~\cite{brink}. By this reason the 
notion of ``spin'' carried by a state is obscured and can be clarified
only after explicit evaluation of string scattering amplitudes~\cite{ov}.
   
Recently, Berkovits and Vafa introduced a new topological theory based on 
a small $N=4$ SCA~\cite{berkvafa} (a similar 
extension of $N=2$ SCA has been considered earlier by
Siegel~\cite{ws}). After topological twisting,
which incidentally does not treat all the fermionic currents symmetrically 
and breaks $SO(2,2)$ down to $U(1,1)$~\cite{berkvafa},
the $N=4$ formalism turns out to be equivalent to the conventional
$N=2$ formalism. This has been demonstrated by 
explicit evaluation of scattering amplitudes~\cite{berkvafa}. It was 
revealed also that the topological prescription for calculating superstring
amplitudes is free of total--derivative ambiguities (for further developments
see Ref.~\cite{bvw}). It is noteworthy, that a global automorphism
group of the small $N=4$ SCA contains the full Lorentz group 
$SO(2,2)\simeq SU(1,1)\times SU(1,1)'$ which is larger than  
$U(1,1)\simeq U(1)\times SU(1,1)$ intrinsic to $N=2$ SCA.

In the present paper we study classical N=2 string within the framework 
of the N=4 topological prescription. An action functional adequate for the
topological formalism is constructed, this revealing the full Lorentz 
invariance in the target space. The motivation for this work is two--fold. 
Firstly, it seems reasonable to use the $N=4$ topological action 
for restoring the Lorenz invariance missing in the previous analysis 
on the $N=2$ string scattering amplitudes. Secondly, being reduced to a 
chiral sector, the action is appropriate for describing a chiral half of 
a recently proposed $N=2$ heterotic string with manifest space--time 
supersymmetry~\cite{bgl}. 

The organization of the work is as follows. In the next section we
outline the main features concerning a conventional Lagrangian 
formulation for the $N=2$ string. An extension by topological currents 
and the issue of the classical equivalence are addressed in Sect. 3.
The global automorphism group of the $N=4$ SCA is discussed in Sect. 4
which will give us a key to the construction of the action.
In Sect. 5 we apply Noether procedure to install an extra $U(1,1)$
global invariance in the conventional $N=2$ string action, 
thus raising the global symmetry to the full Lorentz group. Beautifully 
enough, the world--sheet gauge fields fall in a multiplet of 
an $N=4$ $d=2$ supergravity.  Sect. 6 contains Hamiltonian analysis for 
the model at hand which proves to be a necessary complement to that 
in Sect. 5.  We summarize our results and discuss
possible further developments in Sect. 7. Our notations are gathered in 
Appendix. 

\vskip 0.5cm

\noindent
{\bf 2. Classical N=2 string in the Lagrangian framework}\\[-4pt]

\noindent
\vskip 0.4cm

As originally formulated in Ref.~\cite{brink}, the action of the
classical $N=2$ string describes a coupling of an $N=2$ world--sheet 
supergravity (containing a graviton ${e_n}^\a$ ($n$ stands for a flat index), 
a complex gravitino $\chi_{A \a}$, $A=1,2$, and a real vector field $A_\a$) 
to a complex (on-shell) matter multiplet. The latter involves a 
complex scalar $z^n$ and a complex Dirac spinor $\psi_A^n$, with the target   
index $n$ taking values $n=0,1$ in the critical dimension. The Brink--Schwarz
action reads
\bea
&& S_0=-{\textstyle{\frac {1}{2\pi}}}
\int \!\! d \tau d \s \sqrt{-g}
\{ g^{\a\b} \partial_\a z \partial_\b {\bar z} 
-i{\bar \psi} \g^n \partial_\a \psi {e_n}^\a 
 + i \partial_\a {\bar\psi} \g^n \psi {e_n}^\a  +
 {\bar\psi} \g^n \psi A_\a {e_n}^\a +\nonumber\\[2pt]
&& \qquad \qquad (\partial_\a z -
{\textstyle{\frac 12}} {\bar\c}_\a \psi) {\bar \psi} \g^n \g^m \c_\b 
{e_n}^\b {e_m}^\a + (\partial_\a {\bar z} -
{\textstyle{\frac 12}} {\bar\psi} \c_\a) {\bar \c}_\b \g^n \g^m \psi 
{e_n}^\a {e_m}^\b \},
\eea
which explicitly lacks manifest Lorentz covariance. Due to the complex 
structure intrinsic to the formalism, the (spin cover of
the) full target--space Lorentz group $Spin(2,2)=SU(1,1)\times SU(1,1)'$ 
turns out to be broken down to $U(1)\times SU(1,1)\simeq U(1,1)$
which is the global symmetry group of the formalism.

Since in two space--time dimensions (on the world--sheet) all irreducible 
representations of the Lorentz group are one-dimensional, it seems 
convenient to get rid of the
$\g$--matrices. This, in particular, allows one to avoid the extensive use of
cumbersome Fierz identities in checking local symmetries of the action.
Besides, the study of various heterotic theories 
taking a chiral half from the $N=2$ string~\cite{deboer,bgl} 
inevitably appeals to an action of such a type. For the case at 
hand this yields (our notations are gathered in Appendix) 
\bea\label{bs}
&& S_0=-{\textstyle{\frac {1}{2\pi}}}
\int \!\! d \tau d \s \sqrt{-g}
\{ g^{\a\b} \partial_\a z \partial_\b {\bar z} 
+i\sqrt{2} (\psi_{(+)}\partial_\a {\bar \psi}_{(+)}+ 
{\bar \psi}_{(+)}\partial_\a \psi_{(+)}){e_{-}}^\a+
\nonumber\\[2pt]
&& \quad  i\sqrt{2} (\psi_{(-)}\partial_\a {\bar \psi}_{(-)}
+{\bar \psi}_{(-)}\partial_\a \psi_{(-)}){e_{+}}^\a
-\sqrt{2} {\bar \psi}_{(+)}\psi_{(+)} A_\a {e_{-}}^\a -
\sqrt{2} {\bar \psi}_{(-)}\psi_{(-)} A_\a {e_{+}}^\a  +
\nonumber\\[2pt]
&& \quad 2i \partial_\a z {\bar\psi}_{(-)} 
\c_{\b (+)}{e_{-}}^\a {e_{+}}^\b
-2i \partial_\a z {\bar\psi}_{(+)} \c_{\b (-)}{e_{+}}^\a {e_{-}}^\b-
2i \partial_\a {\bar z} {\psi}_{(+)} {\bar \c}_{\b (-)}
{e_{+}}^\a {e_{-}}^\b+
\nonumber\\[2pt]
&& \quad 2i \partial_\a {\bar z} {\psi}_{(-)} 
{\bar \c}_{\b (+)}{e_{-}}^\a {e_{+}}^\b-
2{\bar \psi}_{(+)} \psi_{(-)} {\bar \c}_{\a(+)} \c_{\b (-)} 
{e_{+}}^\a {e_{-}}^\b -2{\bar \psi}_{(-)} \psi_{(+)} {\bar \c}_{\a(-)} 
\c_{\b (+)} {e_{+}}^\b {e_{-}}^\a + 
\nonumber\\[2pt]
&& \quad {\bar \psi}_{(+)} \psi_{(+)} {\bar \c}_{\b(-)} \c_{\a (-)}
({e_{+}}^\a {e_{-}}^\b+{e_{+}}^\b {e_{-}}^\a)+
{\bar \psi}_{(-)} \psi_{(-)} {\bar \c}_{\b(+)} \c_{\a (+)}
({e_{+}}^\a {e_{-}}^\b+{e_{+}}^\b {e_{-}}^\a) \}.
\eea
Here the $(\pm)$ indices indicate the chirality of an irreducible spinor
representation with respect to the world--sheet (local) Lorentz group 
(see Appendix). These can also be used to mark  
the right and left moving fermionic modes on 
the world--sheet of the string.

Apart from the usual reparametrization invariance, local Lorentz 
transformations and Weyl symmetry, which allow one to bring locally 
the world--sheet metric to the Minkowskian form, the action~(\ref{bs}) reveals 
two more bosonic transformations with real parameters $a$ and $b$
\be\label{a}
\d A_\a =\partial_\a a, \quad \d \psi_{(\pm)}=-{\textstyle{\frac {i}{2}}}
a \psi_{(\pm)}, \quad \d \c_{(\pm)} =-{\textstyle{\frac {i}{2}}}
a \c_{(\pm)}; 
\ee 
\bea\label{b}
&& \d A_\a =e^{-1} \e_{\a\b} g^{\b\g} \partial_\g b, 
\quad \d \psi_{(+)}=-{\textstyle{\frac {i}{2}}}
b \psi_{(+)}, \quad \d \psi_{(-)}={\textstyle{\frac {i}{2}}}
b \psi_{(-)}, \nonumber\\[2pt] 
&& \d \c_{(+)}={\textstyle{\frac {i}{2}}}
b \c_{(+)}, \quad \d \c_{(-)}=-{\textstyle{\frac {i}{2}}}
b \c_{(-)}, 
\eea 
where $e^{-1}={(det({e_{n}}^\a))}^{-1}=\sqrt{-g}$, these being sufficient
to remove the corresponding gauge field $A_\a$, as well as the super 
Weyl transformation with two complex fermionic parameters $\m_{(\pm)}$
\bea
&& \d \c_{\a(+)}=g_{\a\b} {e_{+}}^\b \m_{(-)}, \quad 
\d \c_{\a(-)}=g_{\a\b} {e_{-}}^\b \m_{(+)}, \nonumber\\[2pt]
&& \d A_\a={\textstyle{\frac {1}{\sqrt{2}}}} g_{\a\b} {e_{+}}^\b {e_{-}}^\g
({\bar \m}_{(+)}\c_{\g(-)}+{\bar\c}_{\g(-)}\m_{(+)})+
\nonumber\\[2pt]
&& \qquad {\textstyle{\frac {1}{\sqrt{2}}}} g_{\a\b} {e_{-}}^\b {e_{+}}^\g
({\bar \m}_{(-)}\c_{\g(+)}+{\bar\c}_{\g(+)}\m_{(-)}),
\eea
and an $N=2$ local world--sheet supersymmetry with fermionic complex parameter
$\e_{(-)}$
\bea
&& \d z =i {\bar \e}_{(-)} \psi_{(+)}, \quad \d \psi_{(-)}=0, \quad 
\d \c_{\a(+)}=0, \quad \d {e_{+}}^\a=0,\nonumber\\
&& \d \c_{\a(-)}=\partial_\a \e_{(-)}
+{\textstyle{\frac {i}{2}}}\e_{(-)}A_\a +{\textstyle{\frac {i}{2\sqrt{2}}}}
({\bar\e}_{(-)} \c_{\a(-)}+
\e_{(-)} {\bar\c}_{\a(-)})\c_{\g(-)}{e_{+}}^\g+
{\textstyle{\frac {i}{\sqrt{2}}}}\e_{(-)} \c_{\a(-)} {\bar\c}_{\g(-)} 
{e_{+}}^\g,
\nonumber\\[2pt]
&& \d \psi_{(+)}={\textstyle{\frac {1}{\sqrt{2}}}} \e_{(-)} \partial_\a 
z {e_{+}}^\a  -{\textstyle{\frac {i}{\sqrt{2}}}}\e_{(-)}\psi_{(-)}
{\bar\c}_{\g(+)}{e_{+}}^\g
+{\textstyle{\frac {i}{2\sqrt{2}}}}\psi_{(+)}({\bar\e}_{(-)}\c_{\g(-)}-
\e_{(-)} {\bar\c}_{\g(-)}) {e_{+}}^\g,
\nonumber\\[2pt]
&& \d {e_{-}}^\a=
-{\textstyle{\frac {i}{\sqrt{2}}}} {e_{+}}^\a
({\bar\e}_{(-)}\c_{\g(-)}+
\e_{(-)} {\bar\c}_{\g(-)}){e_{-}}^\g,\nonumber\\[2pt]
&& \d A_\a = {\textstyle{\frac {1}{\sqrt{2}}}} \{ ({\bar\e}_{(-)} 
(\nabla_\g \c_{\a(-)}-\nabla_\a \c_{\g(-)})-
\e_{(-)}(\nabla_\g {\bar\c}_{\a(-)}-\nabla_\a {\bar\c}_{\g(-)}) \} 
{e_{+}}^\g+\nonumber\\[2pt]
&& \qquad \qquad {\textstyle{\frac {i}{2\sqrt{2}}}} \{ A_\g (
\e_{(-)} {\bar\c}_{\a(-)}+{\bar\e}_{(-)}\c_{\a(-)})-
A_\a (\e_{(-)} {\bar\c}_{\g(-)}+{\bar\e}_{(-)}\c_{\g(-)})\}{e_{+}}^\g
-\nonumber\\[2pt]
&& \qquad \qquad {\textstyle{\frac {i}{4}}}
(\e_{(-)} {\bar\c}_{\a(-)}+{\bar\e}_{(-)}\c_{\a(-)})  
{\bar\c}_{\g(-)} \c_{\d(-)} {e_{+}}^\g {e_{+}}^\d,
\eea
and $\e_{(+)}$
\bea
&& \d z =i {\bar \e}_{(+)} \psi_{(-)}, \quad \d \psi_{(+)}=0, \quad  
\d \c_{\a(-)}=0, \quad \d {e_{-}}^\a=0, \nonumber\\[2pt]
&& \d \c_{\a(+)}=-\partial_\a \e_{(+)}
-{\textstyle{\frac {i}{2}}}\e_{(+)}A_\a -{\textstyle{\frac {i}{2\sqrt{2}}}}
({\bar\e}_{(+)} \c_{\a(+)}+
\e_{(+)} {\bar\c}_{\a(+)})\c_{\g(+)}{e_{-}}^\g-
{\textstyle{\frac {i}{\sqrt{2}}}}\e_{(+)} \c_{\a(+)} {\bar\c}_{\g(+)}
{e_{-}}^\g,
\nonumber\\[2pt]
&& \d \psi_{(-)}={\textstyle{\frac {1}{\sqrt{2}}}} \e_{(+)} \partial_\a 
z {e_{-}}^\a  +{\textstyle{\frac {i}{\sqrt{2}}}}\e_{(+)}\psi_{(+)} 
{\bar\c}_{\g(-)}{e_{-}}^\g
-{\textstyle{\frac {i}{2\sqrt{2}}}}\psi_{(-)}({\bar\e}_{(+)}\c_{\g(+)}-
\e_{(+)} {\bar\c}_{\g(+)}) {e_{-}}^\g,
\nonumber\\[2pt]
&& \d {e_{+}}^\a=
{\textstyle{\frac {i}{\sqrt{2}}}} {e_{-}}^\a
({\bar\e}_{(+)}\c_{\g(+)}+
\e_{(+)} {\bar\c}_{\g(+)}){e_{+}}^\g,\nonumber\\[2pt]
&& \d A_\g = -{\textstyle{\frac {1}{\sqrt{2}}}} \{ ({\bar\e}_{(+)} 
(\nabla_\a \c_{\g(+)}-\nabla_\g \c_{\a(+)})-
\e_{(+)}(\nabla_\a {\bar\c}_{\g(+)}-\nabla_\g {\bar\c}_{\a(+)}) \} 
{e_{-}}^\a-\nonumber\\[2pt]
&& \qquad \qquad {\textstyle{\frac {i}{2\sqrt{2}}}} \{ A_\a (
\e_{(+)} {\bar\c}_{\g(+)}+{\bar\e}_{(+)}\c_{\g(+)})-
A_\g (\e_{(+)} {\bar\c}_{\a(+)}+{\bar\e}_{(+)}\c_{\a(+)})\}{e_{-}}^\a
+\nonumber\\[2pt]
&& \qquad \qquad {\textstyle{\frac {i}{4}}}
(\e_{(+)} {\bar\c}_{\g(+)}+{\bar\e}_{(+)}\c_{\g(+)})  
{\bar\c}_{\b(+)} \c_{\d(+)} {e_{-}}^\b {e_{-}}^\d.
\eea
Altogether these allow one to gauge away the gravitino fields $\c_{\a(\pm)}$.
In the equations above $\nabla_\a$ stands for the conventional world sheet 
covariant derivative $\nabla_\a B_\b=\partial_\a B -\Gamma_{\a\b}^\g B_\g$. 

After the gauge fixing (for a more rigorous analysis see Sect. 6)
\be
{e_{+}}^0={e_{-}}^0={e_{+}}^1=-{e_{-}}^1={\textstyle{\frac {1}{\sqrt{2}}}},
\quad A_\a=0,\quad \c_{\a(\pm)}=0,
\ee
only the first three terms survive in the action~(\ref{bs}) and one has to 
keep track of the $N=2$ superconformal currents which arise
from a variation of the action 
with respect to the world--sheet supergravity fields one had before fixing 
the gauge. We write these explicitly in the Hamiltonian form (see
also Sect. 6)
\bea\label{n2}
&&T=(p_z +{\textstyle{\frac {1}{2\pi}}} \partial_1 \bz)
(p_{\bz} +{\textstyle{\frac {1}{2\pi}}} \partial_1 z)-
{\textstyle{\frac {i}{2{\pi}^2}}}(\p_{(+)} \partial_1 {\bp}_{(+)} +
{\bp}_{(+)} \partial_1 \p_{(+)})=0,\nonumber\\[2pt]
&& G=(p_z +{\textstyle{\frac {1}{2\pi}}} \partial_1 \bz) \p_{(+)} =0,
\qquad \bar G=(p_{\bz} +{\textstyle{\frac {1}{2\pi}}} 
\partial_1 z) {\bp}_{(+)} =0,\nonumber\\[2pt]
&& J ={\bp}_{(+)} \p_{(+)} =0,     
\eea
where $(z^n,p_{z n})$, $n=0,1$, form a canonical pair,
$({\bar z}^n, p_{\bar z  n})$, are complex conjugates and
$\psi^n, {\bar\psi}^n$ are a couple of complex conjugate fermions 
\bea\label{brac}
&& \{ z^n (\s), p_{z}^m (\s') \}=\eta^{nm} \delta(\s- \s'), \quad
\{ {\bar z}^n (\s), p_{\bar z}^m (\s') \}=\eta^{nm} \delta(\s- \s'), 
\nonumber\\[2pt]
&&\{ \p_{(+)}^n (\s), {\bp}_{(+)}^m (\s') \}=i \pi \eta^{nm} \delta(\s- \s'),
\eea
with $\eta^{nm}={\it diag} (-,+)$ the Minkowski metric.
There is also an analogous set where
$(p_z +{\textstyle{\frac {1}{2\pi}}} \partial_1 \bz)$ and $\p_{(+)}$ 
are to be exchanged with
$(p_z -{\textstyle{\frac {1}{2\pi}}} \partial_1 \bz)$, $\p_{(-)}$, these 
for the right movers.

\vskip 0.5cm

\noindent

{\bf 3. Adding topological currents }\\[-4pt]

\noindent

\vskip 0.4cm

It is worth mentioning further that the $N{=}2$ superconformal currents 
given in Eq. (\ref{n2}) above are not the maximal closed set one can 
realize on the space of the matter fields. As was pointed out in 
Refs.~\cite{ws,berkvafa}, two more bosonic currents   
\be\label{top1}
\epsilon^{nm} \p_{n (+)} \p_{m(+)} =0, \quad \epsilon^{nm} {\bp}_{n(+)} 
{\bp}_{m(+)} =0,
\ee 
and two more fermionic ones
\be\label{top2}
\epsilon^{nm} (p_{\bz n} +{\textstyle{\frac {1}{2\pi}}} \partial_1 z_n)
\p_{m(+)} =0, \quad \epsilon^{nm} (p_{z n} +{\textstyle{\frac {1}{2\pi}}} 
\partial_1 {\bz}_n) {\bp}_{m(+)} =0,
\ee
extend the algebra up to a small $N{=}4$ SCA. After a topological twist, 
which does not treat all the fermionic currents symmetrically and breaks
$SO(2,2)$ down to $U(1,1)$~\cite{berkvafa},
the $N{=}4$ extension turns out to 
be equivalent to the $N{=}2$ formalism. This has been demonstrated by 
explicit computation of scattering amplitudes~\cite{berkvafa}.
Guided by the quantum equivalence, it seems quite natural to expect that 
a similar correspondence should hold also at the classical level. 
We now proceed to show that the extra currents~(\ref{top1}),(\ref{top2})
do not contain an additional information as compared to that implied by
the $N=2$ currents. Since equations of motion in both the cases are 
free and have the same form, this provides the classical equivalence.  
It seems convenient first to break a 
target space vector 
index into the light--cone components (see Eq.~(A.3) of Appendix).
Denoting $\Pi_z=(p_z +{\textstyle{\frac {1}{2\pi}}} \partial_1 \bz)$ 
and $\Pi_{\bar z}=(p_{\bz} +{\textstyle{\frac {1}{2\pi}}} \partial_1 z)$,
one finds for the currents~(\ref{n2}) (for brevity we omit spinor 
indices carried by the fermions)
\bea\label{lc}
&& T=-\Pi_z^{+} \Pi_{\bar z}^{-}  -\Pi_z^{-} \Pi_{\bar z}^{+}
+{\textstyle{\frac {i}{2{\pi}^2}}}(\p^{+} \partial_1 {\bp}^{-} +
\p^{-} \partial_1 {\bp}^{+} + {\bp}^{+} \partial_1 \p^{-}
+{\bp}^{-} \partial_1 \p^{+})=0,\nonumber\\[2pt]
&& G=-\Pi_z^{+}\p^{-}-\Pi_z^{-}\p^{+}=0,
\quad \bar G=-\Pi_{\bar z}^{+} {\bar\p}^{-}-\Pi_{\bar z}^{-} {\bar\p}^{+}=0,
\nonumber\\[2pt]
&& J =-{\bp}^{+} \p^{-} -{\bp}^{-} \p^{+} =0,     
\eea
while the newly introduces ones acquire the form
\bea\label{lc1}
&& \p^{+}\p^{-}=0, \quad {\bp}^{+} {\bp}^{-}=0,\nonumber\\[2pt]
&& \Pi_{\bar z}^{+} \p^{-} -\Pi_{\bar z}^{-} \p^{+}=0,
\quad  \Pi_z^{+} {\bp}^{-} -\Pi_z^{-} {\bp}^{+}=0.
\eea
Assuming now that $\Pi_z^{+} \ne 0$ \footnote{This can be achieved, for
example, by imposing a light--cone gauge $z^{+} + {\bar z}^{+} = 
z_0^{+} + {\bar z}_0^{+} +\tau (p_z^{+} +p_{\bar z}^{+})$. The real part of
$\Pi_z^{+}$ (and hence the modulus) then does not vanish and one can safely
divide by  $\Pi_z^{+}$. It should be stressed that, due to the presence of two
temporal dimensions in the target, the light cone analysis
for the $N=2$ string seems to be less rigorous than in the Minkowski space.}
one can readily solve for $\p^{-}$, ${\bp}^{-}$
by making use of the second line of Eq.~(\ref{lc})
\be\label{solutions}
\p^{-}=-{\textstyle{\frac {\Pi_z^{-}}{\Pi_z^{+}}}} \p^{+}, \quad
{\bp}^{-}=-{\textstyle{\frac {\Pi_{\bar z}^{-}}{\Pi_{\bar z}^{+}}}} {\bp}^{+}.
\ee
Being substituted in the expression for $J$ these yield
\be\label{tech}
(\Pi_z^{+} \Pi_{\bar z}^{-}  +\Pi_z^{-} \Pi_{\bar z}^{+}){\bp}^{+}\p^{+}=0.
\ee
It is trivial to observe now that the bosonic currents
from Eq.~(\ref{lc1}) follow from Eq.~(\ref{solutions}). The same turns out 
to be true for the fermionic ones, provided Eq.~(\ref{tech}), 
$T\p^{+}=0$ and $T{\bar\p}^{+}=0$  have been used.

Thus, one can conclude that at the classical level the extra currents of 
the $N=4$ topological description do not contain any new information. 
Our analysis here is in agreement with that of Siegel~\cite{ws}. At the 
quantum level, as an alternative to the proof by Berkovits and Vafa
which appeals to the explicit evaluation of string 
amplitudes~\cite{berkvafa},
one could proceed directly from 
the small $N{=}4$ SCA to verify that the positive 
(half--integer) modes of the fermionic currents (\ref{top2}) 
kill all physical states, provided so do the zero modes 
of bosonic ones~(\ref{top1}). Since even for the smaller $N{=}2$
SCA a proper analysis shows the absence of excited 
states~\cite{bienk} and because the zero modes of the new bosonic currents
annihilate the ground state, one arrives at the same spectrum 
as for the ordinary $N{=}2$ string. 

\vskip 0.5cm

\noindent

{\bf 4. $U(1,1)$ and ${U(1,1)}_{outer}$ }\\[-4pt]

\noindent

\vskip 0.4cm

As has been mentioned above, after the topological twist 
the $N{=}4$ prescription is equivalent to the $N{=}2$ formalism. 
There is an intimate connection between this point and 
automorphism groups corresponding to the algebras. Since this will give
us a key to construction of the action, we turn to discuss this
issue in more detail.

As we have seen earlier, the global symmetry group of the conventional
$N=2$ string, which is also a trivial automorphism of an $N=2$ SCA, 
is given by $U(1,1)$. It seems instructive to give here the explicit
realization. Combining $z^n$ and 
${\bar z}^n$ in a single column
$Z^A =\left(\begin{array}{cccc} z^n\\ {\bar z}^n 
\end{array}\right)$  with $A{=}1,2,3,4$,
one has for the infinitesimal $U(1,1)$ transformation
\be
\delta Z^A=i\a_i {L_i^A}_B Z^B, \qquad {\bar L_i}^{T} \eta =\eta L_i,
\qquad \eta_{AB}={\textstyle{diag}}(-,+,-,+),
\ee 
where $\a^i$ are four real parameters and $L_i$ form a basis in the
$u(1,1)$ algebra, 
$$
\begin{array}{lll}
&& L_1=\left(\begin{array}{cccc} 
0 & -i & 0 & 0\\
-i & 0 & 0 & 0\\
0 & 0 & 0 & -i\\
0 & 0 & -i & 0\
\end{array}\right), \qquad \qquad 

L_2=\left(\begin{array}{cccc} 
-1 & 0 & 0 & 0\\
0 & -1 & 0 & 0\\
0 & 0 & +1 & 0\\
0 & 0 & 0 & +1\
\end{array}\right), \nonumber\\[2pt]

\vspace{0.5cm} 

&& L_3= 
\left(\begin{array}{cccc} 
-1 & 0 & 0 & 0\\
0 & +1 & 0 & 0\\
0 & 0 & +1 & 0\\
0 & 0 & 0 & -1\
\end{array}\right), \qquad \qquad

L_4= 
\left(\begin{array}{cccc} 
0 & +1 & 0 & 0\\
-1 & 0 & 0 & 0\\
0 & 0 & 0 & -1\\
0 & 0 & +1 & 0\
\end{array}\right). 
\end{array}
\eqno{(18)}
$$
\addtocounter{equation}{1}
Because the matrices are block--diagonal the corresponding
representation of the $u(1,1)$ algebra is obviously decomposed into a 
direct sum and the fields $z^n$ and ${\bar z}^n$ do not get mixed 
under these transformations. 

Beautifully enough, one can find another realization of the 
group on the space of the fields at hand that do mix $z^n$ and 
${\bar z}^n$ (generators are off-diagonal) and, what is more important, 
it automatically generates the currents of the small $N=4$ SCA when 
applied to those of the $N=2$ SCA~\cite{bgl}. We stick with the terminology 
of Ref.~\cite{bvw} and call this ${U(1,1)}_{outer}$. The explicit 
form of the generators is
\newpage
$$
\begin{array}{lll}
&& L_1=\left(\begin{array}{cccc} 
0 & -i & 0 & 0\\
-i & 0 & 0 & 0\\
0 & 0 & 0 & -i\\
0 & 0 & -i & 0\
\end{array}\right), \qquad \qquad 

L_2=\left(\begin{array}{cccc} 
0 & 0 & 0 & -i\\
0 & 0 & -i & 0\\
0 & -i & 0 & 0\\
-i & 0 & 0 & 0\
\end{array}\right), \nonumber\\[2pt]

\vspace{0.5cm} 

&& L_3=\left(\begin{array}{cccc} 
-1 & 0 & 0 & 0\\
0  & -1 & 0 & 0\\
0 & 0 & +1 & 0\\
0 & 0 & 0 & +1\
\end{array}\right), \qquad \qquad

L_4=\left(\begin{array}{cccc} 
0 & 0 & 0 & -1\\
0 & 0 & -1 & 0\\
0 & +1 & 0 & 0\\
+1 & 0 & 0 & 0\
\end{array}\right). 
\end{array}
\eqno{(19)}
$$
\addtocounter{equation}{1}
Notice that a combination of the form ${\bar z}^n \eta_{nm} 
y^m$, which is trivially invariant under the action of  
$U(1,1)$, does not hold invariant under the action of ${U(1,1)}_{outer}$, 
and a more general object like ${\bar Z}^A \eta_{AB} Y^B$ is to be handled 
with in a formalism respecting the latter. This, in particular, causes 
certain difficulties, when trying to construct vertex operators for a 
recent $N=2$ heterotic string with manifest space--time 
supersymmetry~\cite{bgl}.

In the next section we shall install ${U(1,1)}_{outer}$ in the $N=2$ string action
by making use of the Noether procedure. Before concluding this section 
we collect some 
technical points regarding the action of $U(1,1)$ and ${U(1,1)}_{outer}$ on 
the elementary objects like $\psi^n \eta_{nm}{\bar\vf}^m$ and 
$\psi^n \e_{nm} \vf^m$. 

Independently of the statistics of the fields $\p$ and
$\vf$, one finds for the outer group

\vspace{0.3cm}
$$
\begin{array}{lll}
\begin{tabular}{|l|c|c|c|c|c|c|c|c|c|c|c|c|}
\hline		\vphantom{$\displaystyle\int$}
${U(1,1)}_{outer}$  & $\d_{\a_1}$ & $\d_{\a_2}$ & $\d_{\a_3}$ & $\d_{\a_4}$ \\
\hline		\vphantom{$\displaystyle\int$}
$\p {\bar\vf}$ & 0 & $\a_2(\p\e\vf-{\bar\p}\e{\bar\vf})$ 
& 0 & $i\a_4(\p\e\vf+{\bar\p}\e{\bar\vf})$\\
\hline		\vphantom{$\displaystyle\int$}
$\p\e\vf$ & 0 & $\a_2(\p{\bar\vf}-{\bar\p}\vf)$ & $-2i\a_3 (\p\e \vf)$  & 
$-i\a_4(\p\bar\vf -{\bar\p}\vf)$\\
\hline		\vphantom{$\displaystyle\int$}
${\bar\p}\e{\bar\vf}$ & 0 & $-\a_2(\p{\bar\vf}-{\bar\p}\vf)$ &
 2i$\a_3({\bar\p}\e{\bar\vf})$  & 
$-i\a_4(\p\bar\vf -{\bar\p}\vf)$\\
\hline
\end{tabular}
\end{array}
\eqno{(20)}
$$
\addtocounter{equation}{1}

while the variations with respect to the ordinary group prove to be much 
simpler
\vspace{0.3cm}
$$
\begin{array}{lll}
\begin{tabular}{|l|c|c|c|c|c|c|c|c|c|c|c|c|}
\hline		\vphantom{$\displaystyle\int$}
$U(1,1)$  & $\d_{\a_1}$ & $\d_{\a_2}$ & $\d_{\a_3}$ & $\d_{\a_4}$ \\
\hline		\vphantom{$\displaystyle\int$}
$\p {\bar\vf}$ & 0 & 0 & 0 & 0\\
\hline		\vphantom{$\displaystyle\int$}
$\p\e\vf$ & 0 & $-2i\a_2 (\p\e \vf)$  & 0 & 0\\
\hline		\vphantom{$\displaystyle\int$}
${\bar\p}\e{\bar\vf}$ & 0 & 2i$\a_2 ({\bar\p}\e{\bar\vf})$  & 0 & 0\\
\hline
\end{tabular}
\end{array}
\eqno{(21)}
$$
\addtocounter{equation}{1}

\vspace{0.2cm}

When realizing ${U(1,1)}_{outer}$ in the $N=2$ string action,
a successive inspection of the terms entering Eq.~(\ref{bs}) 
simplifies considerably with the use of the tables (20), (21). 


\newpage

\noindent

{\bf 5. Restoring Lorentz invariance}\\[-4pt]

\noindent

\vskip 0.4cm

No work has to be done with the kinetic terms, these prove to be 
invariant with respect to both $U(1,1)$ and ${U(1,1)}_{outer}$. The next 
two contributions involving the vector field $A_\a$ are to be 
accompanied by
\bea\label{s1}
&& S_1=-{\textstyle{\frac {1}{2\pi}}}
\int \!\! d \tau d \s \sqrt{-g}
\{ \sqrt{2} (\p_{(-)} \e \p_{(-)} {e_{+}}^\a+
\p_{(+)} \e \p_{(+)} {e_{-}}^\a)(B_\a +iC_\a)-\nonumber\\[2pt]
&& \qquad \qquad \sqrt{2} ({\bar\p}_{(-)} \e {\bar\p}_{(-)} {e_{+}}^\a+
{\bar\p}_{(+)} \e {\bar\p}_{(+)} {e_{-}}^\a)(B_\a -iC_\a) \},
\eea
where $B_\a$ and $C_\a$ are two new real vector (gauge) fields. The whole
piece is inert under ${U(1,1)}_{outer}$ provided the following transformation rules 
for the gauge fields

\vspace{0.3cm}
$$
\begin{array}{lll}\label{19}
\begin{tabular}{|l|c|c|c|c|c|c|c|c|c|c|c|c|}
\hline		\vphantom{$\displaystyle\int$}
${U(1,1)}_{outer}$  & $\d_{\a_1}$ & $\d_{\a_2}$ & $\d_{\a_3}$ & $\d_{\a_4}$ \\
\hline		\vphantom{$\displaystyle\int$}
$A_\a$ & 0 & $-4\a_2 B_\a$ 
& 0 & $-4\a_4 C_\a$\\
\hline		\vphantom{$\displaystyle\int$}
$B_\a$ & 0 & $-\a_2 A_\a$ & $-2\a_3 C_\a$  & 0\\
\hline		\vphantom{$\displaystyle\int$}
$C_\a$ & 0 & 0 &
 $2\a_3 B_\a$  & 
$-\a_4 A_\a$\\
\hline
\end{tabular}
\end{array}
\eqno{(23)}
$$
\addtocounter{equation}{1}

\vspace{0.3cm}
For the ordinary $U(1,1)$ the triplet of vector fields proves to be invariant 
under $\d_{\a_1}$, $\d_{\a_3}$ and $\d_{\a_4}$, while for
$\d_{\a_2}$ one has to set
\be
\d_{\a_2} A_\a =0, \quad  \d_{\a_2} B_\a =-2\a_2 C_\a, \quad
\d_{\a_2} C_\a =2\a_2 B_\a.
\ee
An important point to notice is that the transformations on the gauge fields
do satisfy the $u(1,1)$ algebra. Furthermore, one infers from Eq.~(23) 
that they transform as a triplet of $SU(1,1)$ subgroup of the full
${U(1,1)}_{outer}$, so that one can set a single field
$A_\a^I=(A_\a,B_\a,C_\a)$, $I=1,2,3$.

Let us now proceed to the terms linear in $\partial z, \partial {\bar z}$. 
In order to make them ${U(1,1)}_{outer}$ invariant it suffices to introduce a 
couple of complex fermions $\m_{\a (+)}$ and $\m_{\a (-)}$ which bring
the following contribution to a full action
\bea\label{s2}
&& S_2=-{\textstyle{\frac {1}{2\pi}}}
\int \!\! d \tau d \s \sqrt{-g}
\{ 2i\partial_\a z \e \p_{(-)} \m_{\b (+)} {e_{-}}^\a {e_{+}}^\b
+2i\partial_\a {\bar z} \e {\bar\p}_{(-)} {\bar\m}_{\b (+)} 
{e_{-}}^\a {e_{+}}^\b- \nonumber\\[2pt]
&& \qquad \qquad 2i\partial_\a z \e \p_{(+)} \m_{\b (-)} {e_{+}}^\a {e_{-}}^\b
-2i\partial_\a {\bar z} \e {\bar\p}_{(+)} {\bar\m}_{\b (-)} 
{e_{+}}^\a {e_{-}}^\b \}.
\eea
These should transform according to the rules
\vspace{0.3cm}
$$
\begin{array}{lll}\label{19}
\begin{tabular}{|l|c|c|c|c|c|c|c|c|c|c|c|c|}
\hline		\vphantom{$\displaystyle\int$}
${U(1,1)}_{outer}$  & $\d_{\a_1}$ & $\d_{\a_2}$ & $\d_{\a_3}$ & $\d_{\a_4}$ \\
\hline		\vphantom{$\displaystyle\int$}
$\m_{\a (+)}$ & 0 & $-\a_2 {(\c -{\bar\c})}_{\a(+)}$ 
& $2i\a_3 \m_{\a(+)}$ & $-i\a_4 {(\c -{\bar\c})}_{\a(+)} $\\
\hline		\vphantom{$\displaystyle\int$}
$\m_{\a(-)}$ & 0 &  $-\a_2 {(\c -{\bar\c})}_{\a(-)}$    
& $2i\a_3 \m_{\a(-)}$  &   $-i\a_4 {(\c -{\bar\c})}_{\a(-)} $ \\
\hline		\vphantom{$\displaystyle\int$}
$\c_{\a (+)}$ & 0 & $-\a_2 {(\m-{\bar\m})}_{\a(+)} $ 
& 0 & $i\a_4 {(\m +{\bar\m})}_{\a(+)} $\\
\hline		\vphantom{$\displaystyle\int$}
$\c_{\a (-)}$ & 0 & $-\a_2 {(\m -{\bar\m})}_{\a(-)}$ 
& 0 & $i\a_4 {(\m +{\bar\m})}_{\a(-)} $\\
\hline
\end{tabular}
\end{array}
\eqno{(26)}
$$
\addtocounter{equation}{1}
\vspace{0.3cm}
The action of $U(1,1)$ proves to be nontrivial only for $\d_{\a_2}$
and looks like
\be
\d_{\a_2} \m_{\a(+)}=2i\a_2 \m_{\a(+)}, \qquad 
\d_{\a_2} \m_{\a(-)}=2i\a_2 \m_{\a(-)}.
\ee
It is noteworthy, that the fermionic fields 
$\c_{\a (\pm)}$,  $\m_{\a (\pm)}$ furnish a two--dimensional complex
representation of ${SU(1,1)}_{outer}$ (to be more precise, the representation 
is realized on a four component ``Majorana'' spinor 
$(\m,\c,{\bar\m},{\bar\c})$; the indices $\a$ and $(\pm)$ hold inert). 
One can again set a single complex fermionic field ${\vf}_{\a (\pm)}^i$, 
i=1,2, this composed of $\c_{\a (\pm)}$, $\m_{\a (\pm)}$ and forming a 
doublet representation of  ${SU(1,1)}_{outer}$. As we shall see below no 
extra fields are necessary to make the full action $U(1,1)$ and 
${U(1,1)}_{outer}$ invariant. One eventually comes to the 
conclusion that on the world--sheet the theory is described by
$({e_n}^\a, A_\a^I,{\vf}_{\a A}^i)$. Beautifully enough, this set
coincides with
the $N=4$, $d=2$ supergravity multiplet (see e.g.~\cite{gates}). Thus,
within the framework of the $N=4$ topological formalism the action
describes a coupling of the $N=4$, $d=2$ supergravity to matter multiplets.

We now turn to discuss the last two terms in Eq.~(\ref{bs}).
Introducing a further amendment to the action
\bea\label{s3}
&& S_3=-{\textstyle{\frac {1}{2\pi}}}
\int \!\! d \tau d \s \sqrt{-g} 
\{ (\p_{(+)}\e \p_{(+)}-{\bar\p}_{(+)}\e 
{\bar\p}_{(+)}){(\m-{\bar\m})}_{\a(-)} 
{(\c+{\bar\c})}_{\b(-)}+ \nonumber\\[2pt]
&& \qquad   
(\p_{(+)}\e \p_{(+)}+{\bar\p}_{(+)}\e 
{\bar\p}_{(+)}){(\m+{\bar\m})}_{\a(-)} 
{(\c+{\bar\c})}_{\b(-)}+
(\p_{(-)}\e \p_{(-)}-{\bar\p}_{(-)}\e {\bar\p}_{(-)})\times
\nonumber\\[2pt]
&& \qquad 
{(\m-{\bar\m})}_{\a(+)} {(\c+{\bar\c})}_{\b(+)} + 
(\p_{(-)}\e \p_{(-)}+{\bar\p}_{(-)}\e 
{\bar\p}_{(-)}){(\m+{\bar\m})}_{\a (+)} 
{(\c+{\bar\c})}_{\b (+)} \} \times\nonumber\\[2pt]
&& \qquad {\textstyle{\frac {1}{4}}} 
({e_{+}}^\a {e_{-}}^\b + {e_{+}}^\b {e_{-}}^\a),
\eea
one can compensate their variations. It remains to discuss 
two terms involving both $\p_{(+)}$ and $\p_{(-)}$ and
entering the fourth line in Eq.~(\ref{bs}). Here the analysis turns out to
be more intricate. The most general form of the compensators is 
\bea\label{s3}
&& S_4=-{\textstyle{\frac {1}{2\pi}}}
\int \!\! d \tau d \s \sqrt{-g} 
\{ (\p_{(-)}\e \p_{(+)}-{\bar\p}_{(-)}\e 
{\bar\p}_{(+)}) \Sigma_{\a\b} +(\p_{(-)}\e \p_{(+)}+{\bar\p}_{(-)}\e 
{\bar\p}_{(+)}) \Omega_{\a\b} +\nonumber\\[2pt]
&& \qquad \qquad 2\p_{(-)}{\bar\p}_{(+)}\Lambda_{\a\b}
+2\p_{(+)}{\bar\p}_{(-)}\Pi_{\a\b} \} {e_{+}}^\a {e_{-}}^\b,
\eea
where  $\Sigma_{\a\b}$, $\Omega_{\a\b}$, $\Lambda_{\a\b}$ and
$\Pi_{\a\b}$ are some as yet unspecified tensors. In the following,
we shall assume that $\Sigma,\Omega,\Lambda, \Pi \rightarrow 0$ as 
$\m_{(\pm)} \rightarrow 0$. This seems natural because in the 
limit the action we are searching for should reduce to the $N=2$ string 
action. In order that the action be real the compensators  must obey
\be
{(\Sigma_{\a\b})}^{*}=\Sigma_{\a\b}, \quad {(\Omega_{\a\b})}^{*}=
-\Omega_{\a\b},
\quad {(\Lambda_{\a\b})}^{*}=\Pi_{\a\b}.
\ee 
A variation of the whole piece with respect to $U(1,1)$ and 
${U(1,1)}_{outer}$ yields a set of constraints on the newly introduced 
fields. For example, the $\d_{\a_2}$ transformation implies
\bea\label{comp}
&& \d_{\a_2} \Sigma_{\a\b} +2\a_2[\Lambda_{\a\b} +\Pi_{\a\b} +
{\bar\c}_{\a(+)} \c_{\b(-)} +{\bar\c}_{\b(-)} \c_{\a(+)}]=0, \nonumber\\[2pt]
&& \d_{\a_2} \Lambda_{\a\b} +\a_2[\Sigma_{\a\b} -
{\bar\c}_{\a(+)}{(\m -{\bar\m})}_{\b(-)}
-\c_{\b(-)} {(\m -{\bar\m})}_{\a(+)}]=0, \nonumber\\[2pt]
&& \d_{\a_2} \Pi_{\a\b} +\a_2[\Sigma_{\a\b} -
{\bar\c}_{\b(-)}{(\m -{\bar\m})}_{\a(+)}
-\c_{\a(+)} {(\m -{\bar\m})}_{\b(-)}]=0, 
\eea
with $\Omega_{\a\b}$ being inert. With the help of the tables (26) and (27)
one can readily solve the variational equations. Surprisingly enough, the 
global invariance does not fix the compensators uniquely. One encounters the 
following solution
\bea\label{cross}
&& \Sigma_{\a\b}= c_1 {(\c-{\bar\c})}_{\b(-)} {(\m+{\bar\m})}_{\a(+)}+
c_1 {(\c-{\bar\c})}_{\a(+)} {(\m+{\bar\m})}_{\b(-)}+
c_2 {(\c-{\bar\c})}_{\a(-)} {(\m+{\bar\m})}_{\b(+)}+
\nonumber\\[2pt]
&& \qquad c_2 {(\c-{\bar\c})}_{\b(+)} {(\m+{\bar\m})}_{\a(-)}+
\c_{\a(+)} \m_{\b(-)}- {\bar\c}_{\a(+)} {\bar\m}_{\b(-)}+
\c_{\b(-)} \m_{\a(+)}-{\bar\c}_{\b(-)} {\bar\m}_{\a(+)},
\nonumber\\[2pt]
&& \Omega_{\a\b}= c_1 {(\c-{\bar\c})}_{\b(-)} {(\m-{\bar\m})}_{\a(+)}+
c_1 {(\c-{\bar\c})}_{\a(+)} {(\m-{\bar\m})}_{\b(-)}+
c_2 {(\c-{\bar\c})}_{\a(-)} {(\m-{\bar\m})}_{\b(+)}+\nonumber\\[2pt]
&& \qquad c_2 {(\c-{\bar\c})}_{\b(+)} {(\m-{\bar\m})}_{\a(-)}+
\c_{\a(+)} \m_{\b(-)} +{\bar\c}_{\a(+)} {\bar\m}_{\b(-)}+
\c_{\b(-)} \m_{\a(+)}+{\bar\c}_{\b(-)} {\bar\m}_{\a(+)},
\nonumber\\[2pt]
&& \Lambda_{\a\b}=c_1(\m_{\a(+)} {\bar\m}_{\b(-)}+
\m_{\b(-)} {\bar\m}_{\a(+)}) +c_2(\m_{\a(-)} {\bar\m}_{\b(+)}
+\m_{\b(+)} {\bar\m}_{\a(-)})-{\bar\m}_{\a(+)}\m_{\b(-)},
\eea
with $c_1$ and $c_2$ being arbitrary real numbers. Obviously, the missing 
information needed to fix them is encoded in local symmetries an ultimate
action must possess. As is clear from our discussion above, one expects 
the action to exhibit $N=4$ local supersymmetry. 

Alternatively, one could proceed to the
Hamiltonian formalism and fix the constants there from the requirement
that only the set of currents generating a small $N=4$ SCA remains
after completion of the Dirac procedure. Extra constrains are not allowed. 
Because the Lagrangian and Hamiltonian 
methods are in one--to--one correspondence, this will yield a correct 
answer. It should be mentioned that, generally,  
it may happen to be an intricate task to reveal all the local symmetries
just from the form of an action at hand. For example, the $b$--symmetry of 
the $N=2$ string action displayed above in Eq.~(\ref{b}) was missing in the 
original paper~\cite{brink} and has been discovered only four years later 
in Ref.~\cite{fradkin}. As is well known, a straightforward way to discover
how many local symmetries one should expect to find in a Lagrangian theory is 
prompted by the Hamiltonian framework. For models with irreducible 
constraints it suffices to count the number of Lagrange multipliers 
corresponding to primary first class constraints (see e.g.~\cite{henn},
\cite{der}). In order to elucidate this point for the model under 
consideration and for the sake of coherence we then proceed to the 
Hamiltonian formalism to fix the missing constant.

\vskip 0.5cm

\noindent

{\bf 6. Hamiltonian analysis}\\[-4pt]

\noindent

\vskip 0.4cm

Introducing momenta associated to the configuration space variables 
one find the following primary constraints 
(we define a  momentum conjugate to a fermionic variable to be the
right derivative of a Lagrangian with respect to velocity)
\bea\label{primary}
p_e=0, \quad p_A =0, \quad p_B =0, \quad p_C =0, \quad p_\c=0,
\quad p_{\bar\c}=0, \quad  p_\m=0, \quad p_{\bar\m}=0,
\eea
and
\bea\label{primary1}
&& \Theta_{\p_{(+)}} \equiv p_\p^{(+)} + {\textstyle{\frac {i\sqrt{2}}
{2\pi e}}} {\bar\psi}_{(+)}
{e_{-}}^0=0, \quad  
\Theta_{{\bar\p}_{(+)}} \equiv
p_{\bar\p}^{(+)} + {\textstyle{\frac {i\sqrt{2}}{2\pi e}}}
\p_{(+)} {e_{-}}^0=0, \nonumber\\[2pt]  
&&  \Theta_{\p_{(-)}} \equiv
p_\p^{(-)} + {\textstyle{\frac {i\sqrt{2}}{2\pi e}}} {\bar\psi}_{(-)}
{e_{+}}^0=0, \quad  
\Theta_{{\bar\p}_{(-)}} \equiv
p_{\bar\p}^{(-)} + {\textstyle{\frac {i\sqrt{2}}{2\pi e}}}
\p_{(-)} {e_{+}}^0=0,   
\eea
where $e=det ({e_n}^\a)$ and $p_q$ stands for a momentum canonically 
conjugate to a configuration space variable $q$.
It is worth mentioning now that, a Hamiltonian of a diffeomorphism invariant 
theory is given by a linear combination of constraints (Dirac's theorem).
Remarkably, the form of 
the coefficients $c_1$ and $c_2$ is uniquely fixed already from the form 
of the Hamiltonian. The key point to notice is that, 
since the currents under consideration do no involve 
$\p_{(+)}$ and $\p_{(-)}$ simultaneously, such terms should disappear from 
the Hamiltonian. This automatically holds in the Hamiltonian of the 
$N=2$ string 
\bea
&& H_{N=2} =\int d \s \{ p_e \l_e + p_A \l_A + p_\c \l_\c
+p_{\bar\c} \l_{\bar\c}+ \Theta_{\p_{(+)}} \l_{\p_{(+)}} +
\Theta_{\p_{(-)}} \l_{\p_{(-)}} + \Theta_{{\bar\p}_{(+)}} 
\l_{{\bar\p}_{(+)}} +\nonumber\\[2pt]
&& \qquad \Theta_{{\bar\p}_{(-)}} \l_{{\bar\p}_{(-)}}+
{\textstyle{\frac {i\sqrt{2}}{2\pi e}}} (\p_{(+)}\partial_1 {\bar\p}_{(+)}
+{\bar\p}_{(+)}\partial_1 \p_{(+)}) {e_{-}}^1+
{\textstyle{\frac {i\sqrt{2}}{2\pi e}}} (\p_{(-)}\partial_1 {\bar\p}_{(-)}
+{\bar\p}_{(-)}\partial_1 \p_{(-)}) {e_{+}}^1+\nonumber\\[2pt]
&& \qquad {\textstyle{\frac {\pi e}{ {e_{+}}^0 {e_{-}}^0 }}} (p_z p_{\bar z} +
{\textstyle{\frac {1}{{(2\pi)}^2}}} \partial_1 z \partial_1 {\bar z})-
{\textstyle{\frac {1}{2 {e_{+}}^0 {e_{-}}^0 }}}( {e_{+}}^0 {e_{-}}^1+
{e_{-}}^0 {e_{+}}^1) (p_z \partial_1 z + p_{\bar z} \partial_1 {\bar z})+
\nonumber\\[2pt]
&& \qquad  i(p_{\bar z} -{\textstyle{\frac {1}{2\pi }}} \partial_1 z) 
{\bar\p}_{(-)} \c_{\b (+)} {\textstyle{\frac {{e_{+}}^\b}{{e_{+}}^0}}}
-i(p_{\bar z} +{\textstyle{\frac {1}{2\pi }}} \partial_1 z) 
{\bar\p}_{(+)} \c_{\b (-)} {\textstyle{\frac {{e_{-}}^\b}{{e_{-}}^0}}}
+i(p_z -{\textstyle{\frac {1}{2\pi }}} \partial_1 {\bar z}) 
\p_{(-)} \times
\nonumber\\[2pt]
&& \qquad {\bar\c}_{\b (+)} {\textstyle{\frac {{e_{+}}^\b}{{e_{+}}^0}}}-
i(p_z +{\textstyle{\frac {1}{2\pi }}} \partial_1 {\bar z}) 
\p_{(+)} {\bar\c}_{\b (-)} {\textstyle{\frac {{e_{-}}^\b}{{e_{-}}^0}}}
-{\textstyle{\frac {\sqrt{2}}{2\pi e }}} {\bar \p}_{(+)} \p_{(+)} A_\a
{{e_{-}}^\a}-{\textstyle{\frac {\sqrt{2}}{2\pi e }}} {\bar \p}_{(-)} \p_{(-)}
A_\a {{e_{+}}^\a} +\nonumber\\[2pt]
&& \qquad {\textstyle{\frac {1}{2\pi e }}} {\bar \p}_{(+)} \p_{(+)}
{\bar \c}_{\b (-)} \c_{\a (-)} ( {e_{+}}^\a {e_{-}}^\b + 
{e_{-}}^\a {e_{+}}^\b -2  {e_{-}}^\a {e_{-}}^\b
{\textstyle{\frac {{e_{+}}^0}{{e_{-}}^0}}})+
{\textstyle{\frac {1}{2\pi e }}} {\bar \p}_{(-)} \p_{(-)}
{\bar \c}_{\b (+)} \c_{\a (+)} \times
\nonumber\\[2pt]
&& \qquad    ( {e_{+}}^\a {e_{-}}^\b + 
{e_{-}}^\a {e_{+}}^\b -2  {e_{+}}^\a {e_{+}}^\b
{\textstyle{\frac {{e_{-}}^0}{{e_{+}}^0}}}).
\eea
For the full Hamiltonian this turns out to be true only if the 
constants take the following specific value
\be
c_1 =-1, \qquad c_2 =0.
\ee
In this case the full Hamiltonian acquires the form
\bea
&& H_{N=4} = H_{N=2} +\int d \s \{ p_B \l_B + p_C \l_C + p_\m \l_\m
+p_{\bar\m} \l_{\bar\m}+
{\textstyle{\frac {\sqrt{2}}{2\pi e }}}(\p_{(-)}\e \p_{(-)} {e_{+}}^\a
+\p_{(+)} \e \p_{(+)} {e_{-}}^\a) \times
\nonumber\\[2pt]
&& \qquad (B_\a +i C_\a) -{\textstyle{\frac {\sqrt{2}}{2\pi e }}}
({\bar\p}_{(-)}\e {\bar\p}_{(-)} {e_{+}}^\a
+{\bar\p}_{(+)} \e {\bar\p}_{(+)} {e_{-}}^\a) (B_\a -i C_\a)
+ i(p_{\bar z} -{\textstyle{\frac {1}{2\pi }}} \partial_1 z) \e
\p_{(-)} \times
\nonumber\\[2pt]
&& \qquad  \m_{\b (+)} {\textstyle{\frac {{e_{+}}^\b}{{e_{+}}^0}}}
-i(p_{\bar z} +{\textstyle{\frac {1}{2\pi }}} \partial_1 z)\e 
\p_{(+)} \m_{\b (-)} {\textstyle{\frac {{e_{-}}^\b}{{e_{-}}^0}}}
+i(p_z -{\textstyle{\frac {1}{2\pi }}} \partial_1 {\bar z}) \e 
{\bar\p}_{(-)} {\bar\m}_{\b (+)} {\textstyle{\frac {{e_{+}}^\b}{{e_{+}}^0}}}-
\nonumber\\[2pt]
&& \qquad i(p_z +{\textstyle{\frac {1}{2\pi }}} \partial_1 {\bar z}) \e
{\bar\p}_{(+)} {\bar\m}_{\b (-)}{\textstyle{\frac {{e_{-}}^\b}{{e_{-}}^0}}}+
{\textstyle{\frac {1}{8\pi e }}} [ (\p_{(+)} \e \p_{(+)}-
{\bar \p}_{(+)} \e {\bar\p}_{(+)}){(\m-{\bar\m})}_{\a(-)}
{(\c +{\bar\c})}_{\b(-)}+\nonumber\\[2pt]
&& \qquad (\p_{(+)} \e \p_{(+)}+
{\bar \p}_{(+)} \e {\bar\p}_{(+)}){(\m+{\bar\m})}_{\a(-)}
{(\c +{\bar\c})}_{\b(-)}+(\p_{(-)} \e \p_{(-)}-
{\bar \p}_{(-)} \e {\bar\p}_{(-)}){(\m-{\bar\m})}_{\a(+)}\times
\nonumber\\[2pt]
&& \qquad  {(\c +{\bar\c})}_{\b(+)}+(\p_{(-)} \e \p_{(-)}+
{\bar \p}_{(-)} \e {\bar\p}_{(-)}){(\m+{\bar\m})}_{\a(+)}
{(\c +{\bar\c})}_{\b(+)}]( {e_{+}}^\a {e_{-}}^\b + 
{e_{-}}^\a {e_{+}}^\b)+\nonumber\\[2pt]
&& \qquad {\textstyle{\frac {1}{\pi e }}}[\p_{(+)} {\bar\p}_{(+)} {\bar\m}_{\a(-)}
\m_{\b(-)}- {\bar\p}_{(+)} \e {\bar\p}_{(+)} {\bar\m}_{\a(-)}
\c_{\b(-)}-\p_{(+)} \e \p_{(+)} \m_{\b(-)} {\bar\c}_{\a(-)}]
{e_{-}}^\a {e_{-}}^\b  {\textstyle{\frac {{e_{+}}^0}{ {e_{-}}^0}}}
+\nonumber\\[2pt]
&& \qquad {\textstyle{\frac {1}{\pi e }}}[\p_{(-)} {\bar\p}_{(-)} 
{\bar\m}_{\a(+)}\m_{\b(+)}- {\bar\p}_{(-)} \e {\bar\p}_{(-)} {\bar\m}_{\a(+)}
\c_{\b(+)}-\p_{(-)} \e \p_{(-)} \m_{\b(+)} {\bar\c}_{\a(+)}]
{e_{+}}^\a {e_{+}}^\b  {\textstyle{\frac {{e_{-}}^0}{ {e_{+}}^0}}},
\eea
which, as we shall shortly see, is indeed a linear combination of the 
constraints. In the relations above $\l$'s stand for the Lagrange 
multipliers associated to the primary constraints. Given the form of 
the constants $c_1$ and $c_2$, the solutions~(\ref{cross}) simplify 
considerably
\bea
&& \Lambda_{\a\b}=-\m_{\a(+)} {\bar\m}_{\b(-)}, \quad
\Sigma_{\a\b}={\bar\c}_{\a(+)} \m_{\b(-)}+ {\bar\c}_{\b(-)} \m_{\a(+)}
-\c_{\a(+)} {\bar\m}_{\b(-)}-\c_{\b(-)} {\bar\m}_{\a(+)},\nonumber\\[2pt]
&& \Pi_{\a\b}={\bar\m}_{\a(+)} \m_{\b(-)}, \quad
\Omega_{\a\b}=\c_{\a(+)} {\bar\m}_{\b(-)}+\c_{\b(-)} {\bar\m}_{\a(+)}
+{\bar\c}_{\a(+)} \m_{\b(-)}+ {\bar\c}_{\b(-)} \m_{\a(+)}.
\eea 
An ultimate form of the $N=4$ topological string has thus been fixed and 
is given by the sum 
\be\label{fullaction}
S=S_0+S_1+S_2+S_3+S_4.
\ee
It has to be stressed that $SO(2,2)$ global invariance missing
in our starting point $S_0$ is restored in the full action $S$.

We now turn to outline the Dirac procedure. That only the currents of a 
small $N=4$ SCA remain as essential constraints on the system provides a 
consistency check for the formalism developed so far. 

Following Dirac's method, one requires primary constraints to be conserved 
in time. In other words, a point constrained to lie on a surface at some
initial instant of time is not allowed to leave the 
latter in the process of time evolution. Being applied to $p_A=0$, $p_B=0$
and $p_C=0$ from  Eq.~(\ref{primary}) this gives
\be\label{secondary}
{\bar\p}_{(+)} \p_{(+)} =0, \qquad \p_{(+)}\e \p_{(+)}=0,
\qquad {\bar\p}_{(+)} \e {\bar\p}_{(+)}=0,
\ee
plus analogous equations where a $''(+)''$ should be exchanged with a 
$''(-)''$. These 
are to be viewed as secondary constraints. In deriving the relations 
above Eq.~(A.5) of Appendix proves to be helpful. Analogously, the fermionic 
primary constraints from  Eq.~(\ref{primary}) induce further relations
\bea\label{secondary1}
&& (p_z +{\textstyle{\frac {1}{2\pi }}} \partial_1 {\bar z}) \p_{(+)}=0,
\qquad
(p_z -{\textstyle{\frac {1}{2\pi }}} \partial_1 {\bar z}) \p_{(-)}=0,
\nonumber\\[2pt]
&& (p_{\bar z} +{\textstyle{\frac {1}{2\pi }}} \partial_1 z) \e \p_{(+)}=0,
\qquad
(p_{\bar z} -{\textstyle{\frac {1}{2\pi }}} \partial_1 z) \e \p_{(-)}=0,
\eea
as well as their complex conjugates. The remaining equation $p_e=0$ 
in~(\ref{primary}) yields two more secondary constraints
\bea\label{secondary2}
&& (p_z -{\textstyle{\frac {1}{2\pi }}} \partial_1 {\bar z})
(p_{\bar z} -{\textstyle{\frac {1}{2\pi }}} \partial_1 z)+
{\textstyle{\frac {i\sqrt{2}}{2{\pi}^2 e }}} (\p_{(-)} \partial_1 
{\bar\p}_{(-)} +{\bar\p}_{(-)} \partial_1 \p_{(-)}){e_{+}}^0  =0, \quad
\nonumber\\[2pt]
&& (p_z +{\textstyle{\frac {1}{2\pi }}} \partial_1 {\bar z})
(p_{\bar z} +{\textstyle{\frac {1}{2\pi }}} \partial_1 z)-
{\textstyle{\frac {i\sqrt{2}}{2{\pi}^2 e }}} (\p_{(+)} \partial_1 
{\bar\p}_{(+)} +{\bar\p}_{(+)} \partial_1 \p_{(+)}) {e_{-}}^0=0. 
\eea
In their turn the $\Theta$--equations from Eq.~(\ref{primary1}) do not lead
to any new constraints but determine the value of the Lagrange multipliers
$\l_{\p (\pm)}$ and $\l_{{\bar\p} (\pm)}$ (indicating the presence of 
second class constraints). These prove to be complicated 
expressions. Of direct relevance to the forthcoming discussion
are some of their consequences which we list below
\bea
&& {\bar\p}_{(+)} \l_{\p_{(+)}} =-{\bar\p}_{(+)} \partial_1 \p_{(+)}
{\textstyle{\frac {{e_{-}}^1}{ {e_{-}}^0}}}, \quad
{\bar\p}_{(-)} \l_{\p_{ (-)}} =-{\bar\p}_{(-)} \partial_1 \p_{(-)}
{\textstyle{\frac {{e_{+}}^1}{ {e_{+}}^0}}}, 
\quad \p_{(+)} \e \l_{\p_{ (+)}} \approx 0,
\nonumber\\[2pt]
&& \p_{(-)} \e \l_{\p_{ (-)}} \approx 0, \quad
(p_{\bar z} -{\textstyle{\frac {1}{2\pi }}} \partial_1 z) \l_{{\bar\p}_{ (-)}}
=-(p_{\bar z} -{\textstyle{\frac {1}{2\pi }}} \partial_1 z) \partial_1 
{\bar\p}_{(-)} {\textstyle{\frac {{e_{+}}^1}{ {e_{+}}^0}}}-
{\textstyle{\frac {\pi e}{\sqrt{2} {e_{+}}^0 {e_{+}}^0  }}} \times
\nonumber\\[2pt]
&& 
(p_{\bar z} -{\textstyle{\frac {1}{2\pi }}} \partial_1 z)  
(p_z -{\textstyle{\frac {1}{2\pi }}} \partial_1 {\bar z}) {\bar\c}_{\b (+)}
{e_{+}}^\b, \quad  
(p_{\bar z} +{\textstyle{\frac {1}{2\pi }}} \partial_1 z) \l_{{\bar\p} (+)}
=-(p_{\bar z} +{\textstyle{\frac {1}{2\pi }}} \partial_1 z) \partial_1 
{\bar\p}_{(+)} {\textstyle{\frac {{e_{-}}^1}{ {e_{-}}^0}}}+
\nonumber\\[2pt]
&& 
{\textstyle{\frac {\pi e}{\sqrt{2} {e_{-}}^0 {e_{-}}^0  }}}
(p_{\bar z} +{\textstyle{\frac {1}{2\pi }}} \partial_1 z)  
(p_z +{\textstyle{\frac {1}{2\pi }}} \partial_1 {\bar z}) {\bar\c}_{\b (-)}
{e_{-}}^\b, \quad 
(p_{\bar z} -{\textstyle{\frac {1}{2\pi }}} \partial_1 z) \e \l_{\p (-)}
=-(p_{\bar z} -{\textstyle{\frac {1}{2\pi }}} \partial_1 z) \times
 \nonumber\\[2pt]
&& \e \partial_1 \p_{(-)} {\textstyle{\frac {{e_{+}}^1}{ {e_{+}}^0}}}+
{\textstyle{\frac {\pi e}{\sqrt{2} {e_{+}}^0 {e_{+}}^0  }}}
(p_{\bar z} -{\textstyle{\frac {1}{2\pi }}} \partial_1 z) 
(p_z -{\textstyle{\frac {1}{2\pi }}} \partial_1 {\bar z}) {\bar\m}_{\b (+)}
{e_{+}}^\b, \qquad 
(p_{\bar z} +{\textstyle{\frac {1}{2\pi }}} \partial_1 z) \e \l_{\p (+)}
=\nonumber\\[2pt]
&& 
-(p_{\bar z} +{\textstyle{\frac {1}{2\pi }}} \partial_1 z) \e \partial_1 
\p_{(+)} {\textstyle{\frac {{e_{-}}^1}{ {e_{-}}^0}}}-
{\textstyle{\frac {\pi e}{\sqrt{2} {e_{-}}^0 {e_{-}}^0  }}}
(p_{\bar z} +{\textstyle{\frac {1}{2\pi }}} \partial_1 z)   
(p_z +{\textstyle{\frac {1}{2\pi }}} \partial_1 {\bar z}) {\bar\m}_{\b (-)}
{e_{-}}^\b,
\eea
plus their complex conjugates (notice that ${(\l_{\p_{ (\pm)}})}^{*}
=\l_{{\bar\p}_ {(\pm)}}$). The symbol $\approx$ stands for an equality up
to a linear combination of the constraints. 

An important next point to check is that the 
secondary constraints~(\ref{secondary}),(\ref{secondary1}),(\ref{secondary2}) 
are preserved in time. It is straightforward, although a bit tedious, 
exercise to verify that this is indeed the case and no tertiary constraints 
arise or Lagrange multipliers get fixed. The Dirac algorithm thus ends here.

To fix the gauge freedom one encounters in solutions of equations of motion
due to the presence of the unspecified Lagrange multipliers, it is customary 
to impose extra gauge conditions. For the case at hand one can choose them 
in the form
\bea
&& {e_{+}}^0={e_{-}}^0={e_{+}}^1=-{e_{-}}^1={\textstyle{\frac {1}{\sqrt{2}}}},
\quad A_\a=0,\quad B_\a=0,\quad C_\a=0,\nonumber\\[2pt] 
&& \c_{\a(\pm)}=0, \quad {\bar\c}_{\a(\pm)}=0, \quad 
\m_{\a(\pm)}=0, \quad {\bar\m}_{\a(\pm)}=0.
\eea
The conservation in time of the gauges indeed specifies the Lagrange 
multipliers 
\be
\l_e=\l_A=\l_B=\l_C=\l_\c=\l_{\bar\c}=\l_\m=\l_{\bar\m}=0,
\ee
and singles out a unique trajectory from the bunch of those connected by 
gauge transformations. Notice also that the complicated expressions for 
$\l_\p$, $\l_{\bar\p}$ simplify considerably in the gauge chosen
\be
\l_{\p_{(+)}}=\partial_1 \p_{(+)}, \quad 
\l_{\p_{(-)}}=-\partial_1 \p_{(-)}, \quad
\l_{{\bar\p}_{(+)}}=\partial_1 {\bar\p}_{(+)}, \quad 
\l_{{\bar\p}_{(-)}}=-\partial_1 {\bar\p}_{(-)}.
\ee
It remains to notice that the second class constraints from 
Eq.~(\ref{primary1}) can be considered to be fulfilled strongly, 
provided a conventional  Dirac bracket has been introduced. This removes
the momenta $p_\p^{(\pm)}$, $p_{\bar\p}^{(\pm)}$, while leads
$\p$, ${\bar\p}$ to obey
\be
\{ \p_{(+)}^n, {\bar\p}_{(+)}^m \}_D =i\pi \eta^{nm}, \quad
\{ \p_{(-)}^n, {\bar\p}_{(-)}^m \}_D =i\pi \eta^{nm}.
\ee 
Finally, the full Hamiltonian boils down to
\bea
&& H_{N=4}^{gauge} =\int d \s \{ 
-{\textstyle{\frac {i}{2\pi }}} (\p_{(+)}\partial_1 {\bar\p}_{(+)}
+{\bar\p}_{(+)}\partial_1 \p_{(+)}) 
+{\textstyle{\frac {i}{2\pi }}} (\p_{(-)}\partial_1 {\bar\p}_{(-)}
+{\bar\p}_{(-)}\partial_1 \p_{(-)}) +\nonumber\\[2pt]
&& \qquad \qquad 2\pi (p_z p_{\bar z} +
{\textstyle{\frac {1}{{(2\pi)}^2}}} \partial_1 z \partial_1 {\bar z}) \},
\eea
which also guarantees the free dynamics for the physical fields
$z$, $\bar z$, $\p$, $\bar\p$ remaining in the question.

Now let us comment on the number of local symmetries one can expect to
reveal within the Lagrangian framework. Restricting ourselves first to the 
case of the conventional $N=2$ string Hamiltonian $H_{N=2}$, one sees 
that, before the gauge fixing, the Lagrange multipliers 
$\l_{{e_{\pm}}^\a}$, $\l_{A\a}$, $\l_{ {\c_{\a (\pm)}}}$, 
$\l_{{{\bar\c}_{\a (\pm)}}}$ are not determined by the Dirac procedure 
and bring degeneracy into solutions of equations of motion. This is known 
to be in a one--to--one correspondence with the presence of local symmetries
in the Lagrangian framework.  One indeed finds two local diffeomorphisms, 
the Weyl symmetry and the local Lorentz symmetry 
corresponding to $\l_{{e_{\pm}}^\a}$. Associated to  $\l_{A\a}$ are the local 
$a$ and $b$ symmetries we considered in Sect. 2. The remaining four complex 
fermionic Lagrange multipliers $\l_{{\c_{\a (\pm)}}}$ correspond to the 
super Weyl transformations (two complex parameters) and the $N=2$ local 
world--sheet supersymmetry (two complex parameters). Turning to the full
Hamiltonian $H_{N=4}$, one concludes that the action~(\ref{fullaction}) 
must exhibit four more bosonic 
symmetries associated to $\l_{B_\a}$ and $\l_{C_\a}$, the corresponding
parameters  presumably forming a triplet with $a$ and $b$ from 
Eqs.~(\ref{a}), (\ref{b}) under the action of ${SU(1,1)}_{outer}$. The 
latter point is 
also prompted by the fact that the corresponding gauge fields $A_\a$, 
$B_\a$, $C_\a$ do fall in the triplet. 
Besides, one expects to reveal a set of fermionic symmetries with four 
complex parameters, these corresponding to $\l_{{\m}_{\a(\pm)}}$.
Because the gauge fields $\c_{\a (\pm)}$, $\m_{\a (\pm)}$ proved to form
a spinor representation of ${SU(1,1)}_{outer}$, one can indeed expect
a doubling of the super Weyl transformations and the $N=2$ local 
supersymmetry intrinsic to the conventional ($U(1,1)$ covariant) formalism.
Although we did not perform the Lagrangian analysis explicitly, 
on the basis of the Hamiltonian analysis outlined above one can conclude 
that the full action~(\ref{fullaction}) must exhibit an 
$N=4$ local supersymmetry.

\vspace{0.5cm}

\noindent
{\bf 7. Conclusion}

\vspace{0.4cm}

Thus, in the present paper we extended the  $N=2$ string
action to the one adequate for the $N=4$ topological prescription
by Berkovits and Vafa. The major advantage of the new formulation
is that the Lorentz invariance holds manifest. The approach proposed 
in this work differs from the previous ones. We neither use gauging of 
global supersymmetry~\cite{brink} nor appeal to dimensional 
reductions~\cite{pernpvn}. Based on a simple observation that an 
automorphism group of the small $N=4$ SCA involves an extra $U(1,1)$ 
we constructed the $N=4$ topological string action just by installing the 
latter in the $N=2$ string. To guarantee the new global invariance
extra world--sheet fields are to be introduces. Remarkably, they
proved to complement the $N=2$ $d=2$ supergravity multiplet to that of
the $N=4$ $d=2$ world--sheet supergravity.

Turning to possible further developments, the obvious point missing in this 
paper is to explicitly reveal extra local symmetries indicated in Sect. 6.
This seems to be a technical point rather than an ideological one. A more
tempting question is whether it is possible to make use of the action for 
a covariant calculation of scattering amplitudes. It is worth mentioning 
also, that the action functional, when reduced to 
a chiral half, describes the right movers of a recent $N=2$ heterotic 
string with manifest space time supersymmetry~\cite{bgl}. Adding a Lagrangian
describing the left movers might give a heterotic action missing in 
Ref.~\cite{bgl} and suggests another interesting problem.

\vspace{0.6cm}

\noindent
{\bf Acknowledgments}\\[-4pt]

\noindent
We would like to thank Nathan Berkovits for useful discussions.\\
This work was supported by the Iniziativa 
Specifica MI12 of the Commissione IV of INFN. 

\vspace{0.6cm}

\noindent
{\bf Appendix}\\[-4pt]

\noindent
It seems customary to use purely imaginary $\gamma$--matrices
to describe spinors on the world sheet of a string. A conventional basis is 
$$
\begin{array}{lll}
\g_0=\left(\begin{array}{cccc} 
0 & -i\\
i & 0\\
\end{array}\right), \qquad 
\g_1=\left(\begin{array}{cccc} 
0 & i\\
i & 0\\
\end{array}\right), \qquad
\g_3=\g_0 \g_1 =\left(\begin{array}{cccc} 
1 & 0\\
0 & -1\\
\end{array}\right).
\end{array}
\eqno{(A.1)}
$$
It is trivial to check the algebraic properties
$$
\begin{array}{lll}
\{ \gamma_m, \gamma_n \} =-2\eta_{mn}, \quad 
\gamma_m \gamma_n =-\eta_{mn} -\epsilon_{mn} \g_3, \quad
 {(\gamma_0 \gamma_m)}^{T}=\g_0 \g_m,
\end{array}
\eqno{(A.2)}
$$
where $\eta_{mn} ={\it diag} (-,+)$ and $\epsilon_{mn}$ is the 2d Levi-Civita 
totally antisymmetric tensor, $\epsilon_{01}=-1$.

Since in $d=2$ irreducible representations of the Lorentz group
are one--dimensional, it is convenient to use the light-cone notation
for vectors and spinors
$$
\begin{array}{lll}
A_{\pm}={\textstyle{\frac {1}{\sqrt 2}}}(A_0 \pm A_1),
\quad A^n B_n = -A_{+} B_{-} -A_{-} B_{+}, \quad
\Psi_A =\left(\begin{array}{cccc} 
\psi_{(+)}\\
\psi_{(-)}\\
\end{array}\right), 
\end{array}
\eqno{(A.3)}
$$
which makes the latter point more transparent. Actually, 
the Lorentz transformation acquires the form
$$
\begin{array}{lll}
\d A_{\pm} =\pm \o A_{\pm}, \quad \d \psi_{(\pm)}=\pm 
\textstyle{\frac 12}\psi_{(\pm)},
\end{array}
\eqno{(A.4)}
$$
and the invariance is kept, for example,  by contracting a ``${+}$'' with 
a ``${-}$'' (one could fairly well contract a ``${(+)}$'' with a ``${(-)}$'' 
or a ``${+}$'' with two ``${(-})$''). In the light cone notation one can get 
rid of $\gamma$--matrices and work explicitly in terms of irreducible 
components of tensors under consideration.

Introducing the zweibein ${e_m}^\a$ on the world sheet,
$\eta_{nm}={e_n}^\a g_{\a\b} {e_m}^\b$, $g^{\a\b}={e_n}^\a \eta^{nm} {e_m}^\b$,
where $\a$ stands for a curved index and $m$ for a flat one, one can easily 
verify the relations 
$$
\begin{array}{lll}
&& g^{\a\b}=-{e_{+}}^\a {e_{-}}^\b -{e_{-}}^\a {e_{+}}^\b, \quad
\epsilon^{mn} {e_m}^\a  {e_n}^\b = 
-{e_{+}}^\a {e_{-}}^\b +{e_{-}}^\a {e_{+}}^\b=e \epsilon^{\a\b},
\nonumber\\[2pt]
&& \eta_{++}=\eta_{--}=0, \qquad \eta_{+-}=\eta_{-+}=-1,
\end{array}
\eqno{(A.5)}
$$
where $e={\it det} ({e_m}^\a)$.

\end{document}